\documentclass{elsart}
\usepackage{amssymb}
\usepackage{epsfig}

\begin{document}

\begin{frontmatter}

\title{Data acquisition system for the MuLan muon lifetime experiment.}

\author[UK]{V.~Tishchenko} 
\author[UK]{S.~Battu} 
\author[UK]{S.~Cheekatmalla} 
\author[UIUC]{D.B.~Chitwood}
\author[UK]{S.~Dhamija}
\author[UK]{T.P.~Gorringe} 
\author[UIUC,UCB]{F.~Gray}\cite{Regis}
\author[BU]{K.R.~Lynch} 
\author[BU]{I.~Logashenko}
\author[UK]{S.~Rath}
\author[UIUC]{D.M.~Webber}

\address[UK]{Department of Physics and Astronomy,
University of Kentucky, Lexington, KY 40506, USA}
\address[UIUC]{Department of Physics,
University of Illinois at Urbana-Champaign, Urbana, IL 61801, USA}
\address[UCB]{Department of Physics,
University of California, Berkeley, CA 94720, USA}
\address[BU]{Department of Physics,
Boston University, Boston, MA 02215, USA}

\begin{abstract}
We describe the data acquisition system
for the MuLan muon lifetime experiment
at Paul Scherrer Institute. The system 
was designed to record muon decays
at rates up to 1~MHz and acquire data at rates up to 
60~MB/sec. The system
employed a parallel network of dual-processor machines 
and repeating acquisition cycles of deadtime-free time segments
in order to reach the design goals. The system incorporated 
a versatile scheme for control and diagnostics
and a custom web interface for monitoring experimental conditions.
\end{abstract}


\end{frontmatter}

\section{Introduction}
\label{s:Introduction}

The MuLan experiment was designed to measure the positive
muon lifetime to a precision of one part-per-million (ppm), 
a twenty-fold improvement over earlier experiments.
The muon lifetime $\tau_{\mu^+}$ 
determines the Fermi constant G$_F$,
which governs the strength of all weak interaction processes. 
In combination with other quantities, a precision determination
of G$_F$ allows for precision testing of the standard model.

The MuLan experiment was located in the $\pi$E3 beamline at
Paul Scherrer Institute. 
It used a custom-built electrostatic kicker \cite{Ba04}
to produce an intense, low-energy, pulsed muon beam with beam-on periods of
typically 5~$\mu$s and beam-off periods of typically 22~$\mu$s. 
During the beam-on (or accumulation) periods, the incoming muons 
were accumulated in a stopping target, and during the beam-off  
(or measurement) periods, the outgoing decay positrons 
were observed in a scintillator array. 
By recording the times of the outgoing positrons during the measurement period 
the $\mu^+$ decay curve was constructed and 
the muon lifetime $\tau_{\mu^+}$ was extracted.  

Reaching a precision of one part-per-million in $\tau_{\mu^+}$
requires both enormous statistics (exceeding 10$^{12}$ decays)
and careful control and comprehensive monitoring of systematic effects. 
To date, we have published results  from our first measurement of the muon
lifetime, with a precision of 11~ppm \cite{Ch07}. The analysis of data collected in
two production runs, each with $10^{12}$ recorded muon decays, is
currently underway.
 
In this paper we describe the design, operation and performance of the data
acquisition system that was developed for the MuLan experiment.
In section \ref{s:Setup}, we briefly describe the experimental
design and measuring devices.
In section \ref{s:Acquisition}, we discuss the design and
layout of the data acquisition system, give details of our
methods for control synchronization and data readout, and discuss the
system performance.  Finally, we discuss the custom system
developed for online control and run monitoring in section \ref{s:Monitoring}. 

\section{MuLan setup}
\label{s:Setup}

\begin{figure}
\begin{center} 
\mbox{\epsfig{figure=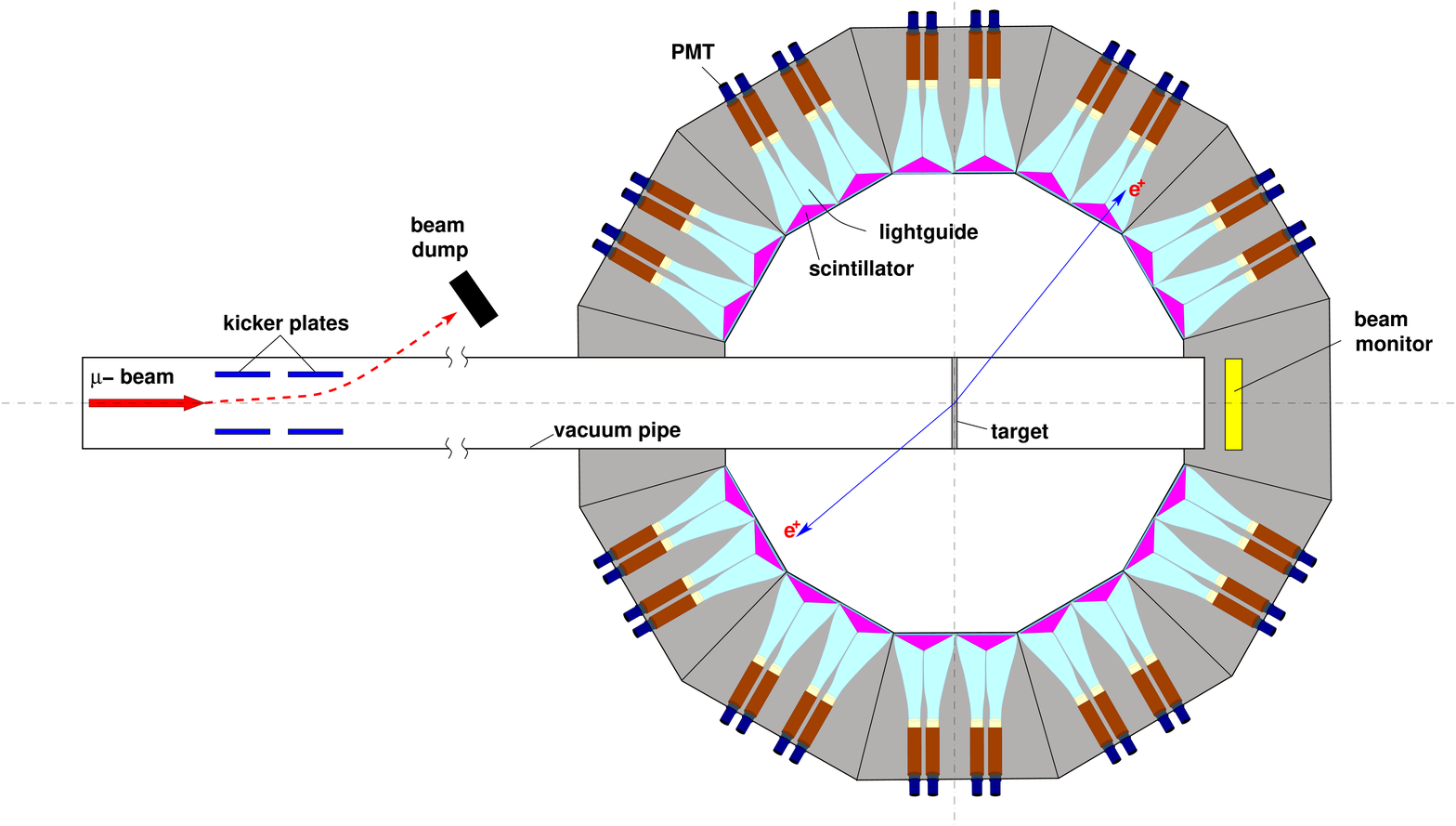,width=0.95\linewidth}}
\end{center}
\caption{Schematic view (not to scale) of the MuLan experimental setup
showing a cross section view of the 
electrostatic kicker, vacuum pipe, target assembly, beam
monitor and plastic scintillator array (MuLan ball).
Note that the stopping targets were
mounted inside the beam pipe vacuum, 
and were rotated into the beam for production running
and out of the beam for beam studies.}
\label{f:schematic}
\end{figure}

A schematic view of the 
MuLan setup is shown in Fig.\ \ref{f:schematic}. 
The setup comprised a target assembly for
accumulating muons, a high-rate wire chamber 
for beam monitoring, and a fast-timing, finely-segmented, large-acceptance 
scintillator array for recording decay positrons. The stopping targets were
mounted inside the beam pipe vacuum and were rotated into the beam for production running
and out of the beam for beam studies. The high-rate wire chamber (denoted the beam monitor) 
-- with one horizontal and one vertical wire plane -- was located downstream 
of the target assembly and immediately following the beam-pipe exit window.
The scintillator array (denoted the MuLan ball) consisted of 170 triangular
plastic scintillator tile pairs that were arranged in the form of a truncated 
icosahedron ({\it i.e.}\ a soccer ball geometry).

\begin{figure}
\begin{center} 
\mbox{\epsfig{figure=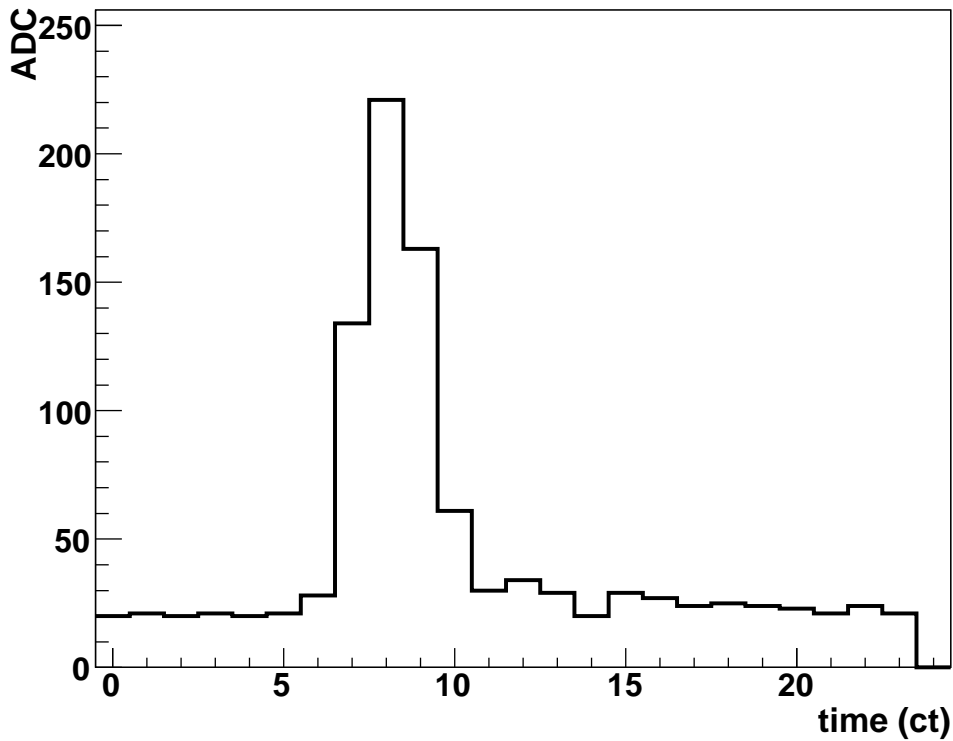,width=0.95\linewidth}}
\end{center}
\caption{Sample pulse recorded by the 500~MHz, 8 bit
waveform digitizers instrumenting the scintillator tile array. 
The horizontal axis is in clock ticks (ct) with $\sim$2~ns per tick.}
\label{f:pulse}
\end{figure}

The $2$$\times$$170$ scintillator tiles were each 
read out via a light pipe and a photomultiplier tube to a \textsc{vme}-based, 
500 MHz, 8~bit waveform digitizer. 
When an input signal exceeded a programmable threshold 
the waveform digitizer wrote both a time stamp and a consecutive sequence of 
flash \textsc{adc} samples to the onboard \textsc{fifo} memory 
of the corresponding digitizer channel (Fig. \ref{f:pulse} shows a
typical digitized pulse). 
The wire chamber was read out via pre-amplifiers and 
discriminators to a custom-built \textsc{fpga} that derived
the x-y coordinates and hit times of the wire-plane coincidences.
All chamber coincidences during the measurement period,
and pre-scaled chamber coincidences during the accumulation period,
were written into a \textsc{vme}-based memory unit (Struck \textsc{sis3600}).
Additional systems were responsible for the control and the monitoring
of the electrostatic kicker, beamline magnets, high-voltage system
and other variables relating to experimental conditions.

\section{High speed data acquisition}
\label{s:Acquisition}

\subsection{Design}
\label{s:Design}

The data acquisition in MuLan 
had to confront a number of challenges.
First, to accumulate $10^{12}$ decays implies 
both very high data rates (up to 50~MB/sec) 
and very large data volumes (up to 100~TB) 
must be handled with small deadtimes,
with any necessary deadtimes being scheduled
to avoid distorting the positron time spectrum.
Second, comprehensive monitoring of experimental variables
and straightforward configuration for diagnostic measurements
must be accommodated.

In MuLan the acquisition does not read out decay positrons one-by-one. 
Rather, to handle the high data rates and avoid 
any deadtime-related distortions,
we collected data in repeating acquisition cycles 
of deadtime-free time segments.\footnote{Under normal
running conditions the waveform digitizer electronics
was gated on during the measurement period
and gated off during the accumulation period 
(see Sec.\ \ref{s:operation} for details).}
The approach utilized the \textsc{vme}-based memories
in the waveform digitizers
to temporarily cache the incoming data.
Each acquisition cycle thus yielded a complete, distortion-free record
of all over-threshold pulse shapes  
from the scintillator tiles 
over the entire duration of the measurement period.
In normal operation the duration of one time segment was 5000 fill cycles
or approximately 135 milliseconds.
Between each time segment a short deadtime of 
roughly 2-4~ms was employed to complete the
transactions that manage the acquisition. 

To handle high rates
and facilitate diagnostic studies,
the data acquisition was designed with a modular, distributed philosophy
and implemented on a parallel, layered processor array. 
A frontend layer of seven dual-processors was used
for the parallel readout of the waveform digitizers
and the beam monitor.
A slow control layer was used
for the control and monitoring
of other instrumentation such as high-voltage supplies
and beamline components.
A backend layer was responsible 
for the assembly of the event fragments 
and the storage of the reconstructed events.\footnote{Herein we use the  
term ``event'' to indicate 
the software assemblies of recorded data that correspond to one time segment
or one acquisition cycle. Typically one event will 
contain many tile pulses and many decay positrons.}
Lastly, an online analysis layer was responsible
for monitoring and diagnostics
that ensured the integrity of the recorded data.

\subsection{Layout}
\label{s:Layout}

\begin{figure}
\begin{center} 
\mbox{\epsfig{figure=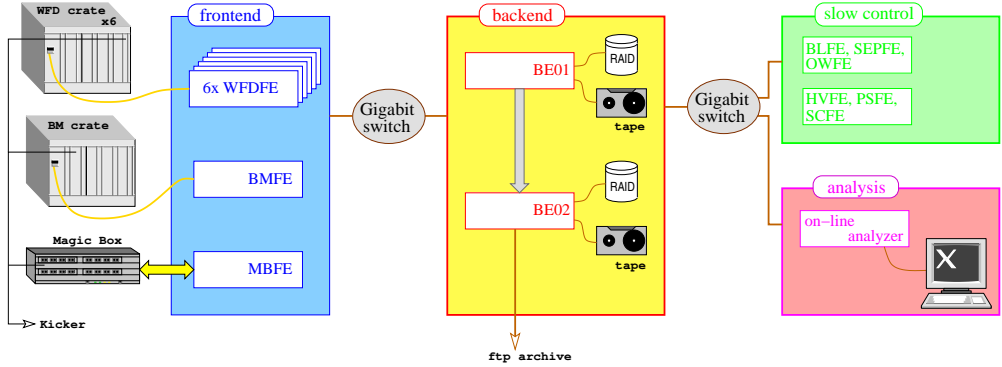,width=0.95\linewidth}}
\end{center}
\caption{Schematic view of the acquisition layout. Shown
are the frontend layer responsible for the waveform digitizer
and beam monitor read out, the backend layer responsible
for event building and data storage, the slow control layer 
responsible for control and diagnostics, and the online analysis layer
responsible for integrity checks and basic histogramming.. 
See Sec.\ 3.2 for the description of the individual components of the
various layers.}
\label{f:layout}
\end{figure}

The data acquisition system is depicted schematically 
in Fig.\ \ref{f:layout}.  It shows the frontend layer 
responsible for waveform digitizer readout and beam monitor readout, 
the slow control layer responsible for control and monitoring,
the backend layer responsible for event assembly and data storage, 
and the online analysis layer responsible for 
basic histogramming and integrity checks.
The hardware platform comprised a networked cluster of Intel 
Xeon 2.6~GHz processors (with 1-4 gigabytes of memory)
running Fedora Core 3 Linux. The acquisition software
was developed using the \textsc{midas} data acquisition \cite{midas}
and \textsc{root} data analysis \cite{root} frameworks.
To eliminate packet collisions that would degrade network
traffic flow, we used two private gigabit ethernet networks.  The
first network handled traffic between the frontend layer and the
backend layer, while the second handled traffic between the backend
layer and the slow control and online analysis layers.  This network
segmentation significantly boosted performance and reduced
the deadtimes between the frontend and backend layers.

A key component of the acquisition system was the so-called
``magic box'' frontend (\textsc{mbfe}) process. This process was
responsible for synchronizing the software defined data acquisition
cycles ({\it i.e.}\ time segments) and the hardware beam on/off cycles 
({\it i.e.}\ fill periods).
To coordinate the software components,
the \textsc{mbfe} process used remote procedure calls (\textsc{rpc}s)
for network communication between data acquisition modules
on different processors. To coordinate the hardware components,
the \textsc{mbfe} operated an \textsc{fpga}-based programmable pulser
that issued command signals to the beam kicker, 
waveform digitizers and beam monitor. 
In this manner all deadtimes were executed between the \textsc{midas} events
rather than during the \textsc{midas} events.
The \textsc{mbfe} process executed
on a dedicated single-processor frontend machine.

The primary source of high-rate data were the 340 channels of waveform digitizers
that record pulses from the 340 scintillator tiles of the MuLan ball.
The digitizers were distributed over six \textsc{vme} crates
and read out by six waveform digitizer frontend 
(\textsc{wfdfe}) processes that executed on six dual-processor frontend machines.
The data were transferred from the \textsc{fifo} memories
of the waveform digitizer  
to the random access memory of the frontend processors 
via Struck \textsc{sis3100/1100} bridges \cite{STRUCK}. Each bridge 
consisted of a \textsc{sis3100} \textsc{vme} card that provided access to each crate's \textsc{vme} bus, a 
\textsc{sis1100} \textsc{pci} card that provided access to each processor's 
\textsc{pci} bus, and a gigabit fiber-optic link 
connecting the two interface cards. 

A further source of high-rate data was the beam monitor wire chamber,
which was generally activated for beam studies
and de-activated for production running. 
The beam monitor data was read out by the
beam monitor frontend (\textsc{bmfe}) process that executed
on an additional dual-processor frontend machine.
The beam monitor data was transferred 
from a Struck \textsc{sis3600} \textsc{vme} memory unit to the frontend processor 
random access memory via an additional Struck \textsc{sis3100/1100} bridge.
The read out of beam monitor data was 
performed continuously during each time segment 
by an asynchronous readout loop, thus preventing any possibility
of memory overflow in the \textsc{sis3600} memory unit.

The various processes for control and monitoring
were executed on two slow control processors
and provided the read out of information such as beamline magnet currents, 
photomultiplier voltages/currents, \textsc{vme} power supply voltages/currents, 
environmental magnetic fields and environmental temperatures.
A beamline frontend (\textsc{blfe}) process 
and separator frontend (\textsc{sepfe}) process 
were responsible for the control
and the read out of the beamline elements.
A high voltage frontend (\textsc{hvfe}) process was
responsible for the control and the read out 
of a LeCroy 1440 multichannel high voltage system
powering the scintillator tiles. 
Additionally, a \textsc{vme}-crate frontend (\textsc{psfe}) process enabled monitoring 
of the \textsc{vme} power supplies and a 1-wire interface frontend (\textsc{owfe}) process  
enabled monitoring of various temperature sensors and  magnetic field
probes. Lastly, a scaler frontend (\textsc{scfe}) process  provided the control and
read out for the \textsc{camac} scalers that record the primary proton current
and beam monitor hits.

Note that the slow control processes read out data
in so-called ``periodic'' mode, {\it i.e.}\ at fixed time intervals
that were asynchronous with the data acquisition 
cycle. Every slow control process wrote one copy of its data as a 
\textsc{midas} databank to be stored in the main data stream and another copy of its data 
as an \textsc{ascii} data file that was used by the online monitoring system.
Note that the \textsc{ascii} text files were written irrespective of whether or not
a \textsc{midas} run was currently underway. 

The data fragments from the frontend processes were 
asynchronously transferred to the backend layer 
across the frontend network.
Initially, the data fragments from the six waveform digitizer processes
and the beam monitor process were transferred to individual memory segments 
on the backend machine \textsc{be01}, {\it i.e.}\ one memory segment 
per frontend process.
The \textsc{midas} event builder (\textsc{mevb}) process then re-assembled 
the fragments from the various \textsc{wfdfe} and \textsc{bmfe} processes 
into complete events. 
The reconstructed events were then written by the 
event builder process to a final memory segment known as the system memory segment. 
Note that the slow control databanks from the various slow control processes 
were directly transferred to the system memory segment and thus 
bypassed event building.

The backend layer used two servers 
and a three stage pipeline to permanently store two copies 
of the entire MuLan dataset.
First, the data were transferred from the system memory segment 
on the backend processor \textsc{be01} to a temporary disk file on a local 
redundant disk array. We used an array of
ten 250~GB disks controlled by a \textsc{pci} \textsc{raid} controller. The \textsc{raid} was
configured as \textsc{raid10} (a nested disk array with striping and mirroring) 
in order to provide both fault tolerance and high \textsc{i/o} performance.
Next, the data files on \textsc{be01} were asynchronously migrated from
disk to both an \textsc{lt03} tape robot system, mounted on processor \textsc{be01},
and a remote redundant disk array, mounted on processor \textsc{be02}.
Finally, the data files on \textsc{be02} were asynchronously migrated to
either a second \textsc{lt03} tape robot system 
or the PSI central data archive. 
This multi-step approach was used to minimize any  possible delays in data-taking 
due to latencies associated with the tape storage 
or the archive storage. In addition, the
disk files on the \textsc{be02}-redundant disk array
were available for any offline analysis work.

The online analysis layer was used for integrity checking and online histogramming.
A dedicated dual-processor machine hosted the 
online analyzer processes responsible for the 
various diagnostic and monitoring tasks.
on the backend processor \textsc{be01}.
The online analyzer received 
events ``as available'' in order to avoid introducing
any delays into the read out and the data storage.
This layer is described in detail in Sec.\ \ref{s:Monitoring}.

\subsection{Frontend synchronization}
\label{s:synchronization}

\begin{figure}
\begin{center} 
\mbox{\epsfig{figure=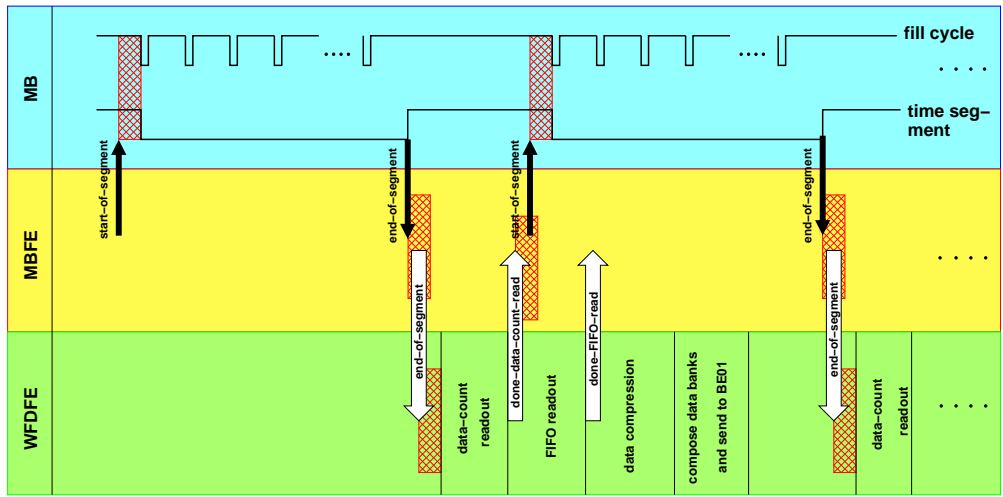,width=0.95\linewidth}}
\end{center}
\caption{Diagram of the synchronization of the fill cycles and the acquisition
cycles across the different components of the data acquisition. The
upper band represents the magic box pulser, the middle band represents
the magic box frontend process, and the lower band represents a
waveform digitizer frontend process. The horizontal traces indicate
the control signals for fill periods and time segments that 
are generated by the pulser. The open arrows represent the 
remote procedure calls between the frontend processes and the filled
arrows represent the parallel port communications with the pulser.
The hashed regions indicate delays associated with either 
the hardware communications or the inter-process communications.
For the waveform digitizer process the 
sequence of data count read out, data read out, data
compression, and databank transfer are also shown.} 
\label{f:operation}
\end{figure}

This section gives detailed information 
on the synchronization of the different components of the data acquisition.
A diagram summarizing the various synchronization signals
that were transmitted between the magic box frontend
and the waveform digitizer frontends is given in
Fig.\ \ref{f:operation}.

To initiate a new time segment the \textsc{mbfe} process sends 
a ``start-of-segment'' message to the magic box programmable pulser 
via a parallel port connection. 
This triggers transmission of a pre-programmed number of 
``run'' gates to the waveform digitizers, 
thus causing each digitizer channel to store the pulses
that were present during these gates. 
Each run gate corresponds to one 22~$\mu$s measurement period
and was synchronized to the beam on-off transitions which were also
controlled by the magic box.

After completing a run gate sequence
the magic box pulser set an ``end-of-segment'' bit.
This bit was identified  by the \textsc{mbfe} process via a polling routine 
and then broadcast to the waveform digitizer frontend processes 
via a remote procedure call. Each 
\textsc{rpc} generated a program interrupt that causes 
the digitizer processes to read (i) the fill count register
for each digitizer module, (ii) the data count register
for each digitizer channel, and (iii) the 
\textsc{fifo} memories of each digitizer channel.\footnote{The 
fill count register stored the total number of run gates 
that were received by the digitizer. 
By reading and clearing this register 
between segments, it accumulated
the number of fills for the previous segment.  The data count register 
stored a pointer to the last entry in the \textsc{fifo} memory of the digitizer channel.
By reading this register between segments it was possible
to determine the number of data words collected during the previous segment.}
Each digitizer process separately reported 
the completion of task (ii), denoted ``done-data-count-read'',
and task (iii), denoted ``done-\textsc{fifo}-read'',
to the \textsc{mbfe} process via an \textsc{rpc}. 
Note that the read out of the digitizer \textsc{fifo} memories could extend into
the following time segment, but the requirement that all ``done-data-count-read''
messages were received before the start of that next segment ensured that
the cached data in the digitizer \textsc{fifo} memories were correctly assigned
to their parent time segment.

After the receipt of all ``done-\textsc{fifo}-read'' \textsc{rpc}s from all 
digitizer processes,
and the identification of the next end-of-segment bit 
in the magic box pulser, the \textsc{mbfe} initiates a new readout cycle 
via a new ``end-of-segment'' \textsc{rpc} to the digitizer 
processes. The ``done-\textsc{fifo}-read'' requirement ensures the
correct sequencing of the \textsc{vme} accesses for the various readout tasks.

\subsection{Digitizer readout}
\label{s:operation}

The waveform digitizer frontend processes performed several tasks 
that included: the read out of the digitizer memories, 
the lossless compression of the digitizer data,
and the assembly of the \textsc{midas} databanks. 
The \textsc{wfdfe} processes ran on dual-processor frontend machines and 
used \textsc{posix} multi-threading functions
to enable the simultaneous execution of multiple 
readout threads (one thread per segment).
Mutexes were used for the thread-unsafe parts of the
frontend code such as reading data from the digitizer memories 
and transferring data to the backend processor.
The \textsc{wfdfe} threads also performed the lossless compression 
of digitizer data using the \textsc{zlib} library.\footnote{\textsc{zlib} is a 
general purpose compression library that is based 
on a combination of the \textsc{lz77} algorithm
and Huffman coding \cite{Hu52}.}
An \textsc{md5} checksum computation by the acquisition
code before data compression and
by the analysis code after data decompression was used
to ensure the integrity of the compressed databank.\footnote{We 
used the \textsc{md5} cryptographic hash function to generate a checksum to
immunize the data stream against errors in compression, decompression,
storage and transmission.  The \textsc{md5} function yielded an efficiently calculable 128 bit
``fingerprint'' of the data. The \textsc{md5} fingerprint was nearly certain to change if any errors 
were introduced into the data stream.}

The waveform digitizer data were stored in
\textsc{midas} data banks with one \textsc{midas} bank per digitizer channel. 
Within each bank the data were formatted as digitizer blocks 
with twenty four consecutive 8-bit \textsc{adc} values, 
a 32-bit alignment block, a 16-bit fill stamp
and a 16-bit time stamp. 
A header block stored the data count 
and fill count that were read from each digitizer 
-- a redundancy enabling additional error checking.

\subsection{Performance}
\label{s:Performance}

In the MuLan experiment, the data was stored on disk and tape in ``runs'',
which comprised time-ordered sequences of software acquisition cycles 
of duration 135~ms, which were further subdivided into hardware-based fill
cycles of duration 27~$\mu$s.  A standard run was acquired over roughly
600 seconds, consumed about 10~GB of storage space, and contained
$4\times10^3$ acquisition cycles or roughly $2\times 10^7$ fill cycles.

Under typical running conditions
the average rate of recorded tile pulses
was about 3~kHz per digitizer channel,
200~kHz per digitizer frontend,
and 1~MHz in total.
We stress these rates were time averages across
the fill cycle, the instantaneous rates 
vary considerably over the fill period.
The above pulse rates were equivalent to approximately 100~kB/s
per digitizer channel, approximately 6~MB/s per digitizer 
frontend, and approximately 36~MB/s in total.
Herein we discuss the rate capabilities of the major components
of the data acquisition -- {\it i.e.}\ 
data read out, data compression, network transfer, event building and event storage --
and the resulting performance of the complete system.

Between time segments -- {\it i.e.}\ between the 
magic box process identifying the end of the previous 
time segment and initiating the start of the following time segment -- the
waveform digitizer processes must read the data count registers
of each digitizer channel and fill count registers of each digitizer module. 
The necessary enhanced parallel-port communications between the magic box process
and the magic box pulser had typical delays of several microseconds.
The necessary remote procedure calls between the magic box process
and the waveform digitizer processes had delays up to 100 microseconds.
By comparison 
the sequence of about 60 data count reads (one read per digitizer channel) 
and about 15 fill count reads (one read per digitizer module) 
across an entire \textsc{vme} crate took several milliseconds -- thereby 
dominating the intervening deadtime between successive segments.
Consequently, the data count and fill count reads imposed a data-rate independent 2-4~ms deadtime 
between successive time segments (a few percent deadtime for 135~ms time segments).

The data volume cached during the time segment determined the time
required to read out and compress the digitizer data.
Below a certain critical data rate the 
read out and compression were completed during 
the acquisition of the next segment, thereby 
adding no additional deadtime between acquisition cycles. 
Above this critical data rate the read out and compression 
were not completed during the acquisition of the next segment,
thereby extending the intervening deadtime between acquisition cycles
(for details see Secs.\ \ref{s:synchronization} and \ref{s:operation}). 
In practice, the critical data rate for minimal deadtime was found
to be approximately 60~MB/s of uncompressed data or 
approximately 40~MB/s of compressed data.

Following the read out and compression, the potential bottlenecks
in data acquisition were the network transfer, event building 
and data storage. Event building involved copying data
between memory segments and event storage involved
copying data from memory segments to disk and tape.
These tasks were limited by the rate capabilities
of \textsc{i/o} operations, both to and from memory, disk and tape. 
If event building was unable to keep pace, it blocked the data transfer from the
frontend processes, thus potentially inhibiting the data acquisition.
If data logging was unable to keep pace, 
it blocked the data transfer from the event builder, 
again potentially inhibiting the data acquisition. 
Under normal conditions  of about 25~MB/s of compressed data the
acquisition performance was not limited by network transfers, event
building and data storage rates.
However, for rates of 35~MB/sec and higher the system 
was unable to maintain the storage of two copies of the data-set. 
Note that data compression in the digitizer frontends
--  which reduced the data rates through event building
and data storage by roughly 30\% -- was important
in circumventing the rate limitations
of network transfer, event building and data storage 
for high rate operation.

\begin{figure}
\begin{center} 
\mbox{\epsfig{figure=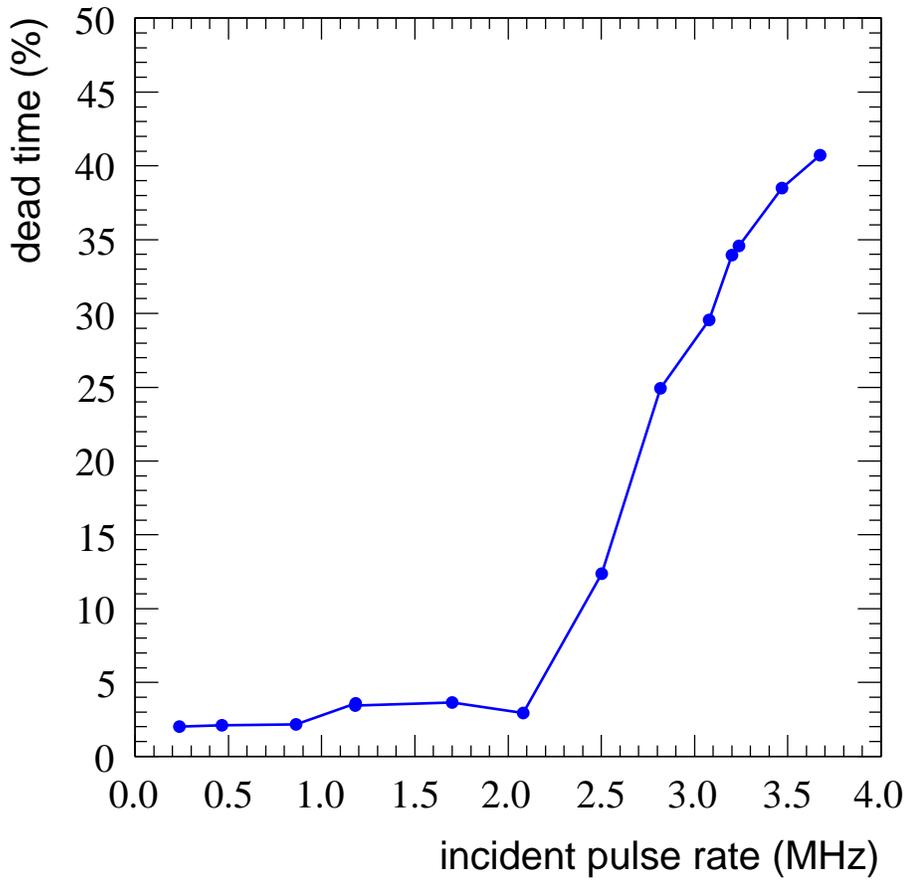,width=0.95\linewidth}}
\end{center}
\caption{The deadtime versus incident pulse rate for the
full data acquisition setup including digitizer read out, lossless compression,
event building and data storage. At pulse rates below 2~MHz 
the deadtime is several percent and
dominated by the read out of the data count registers and the fill count registers
in the waveform digitizers. 
At pulse rates above 2~MHz the deadtime is linearly increasing 
and dominated by the read out and the compression of the digitized pulses.} 
\label{f:performance1}
\end{figure}

\begin{figure}
\begin{center} 
\mbox{\epsfig{figure=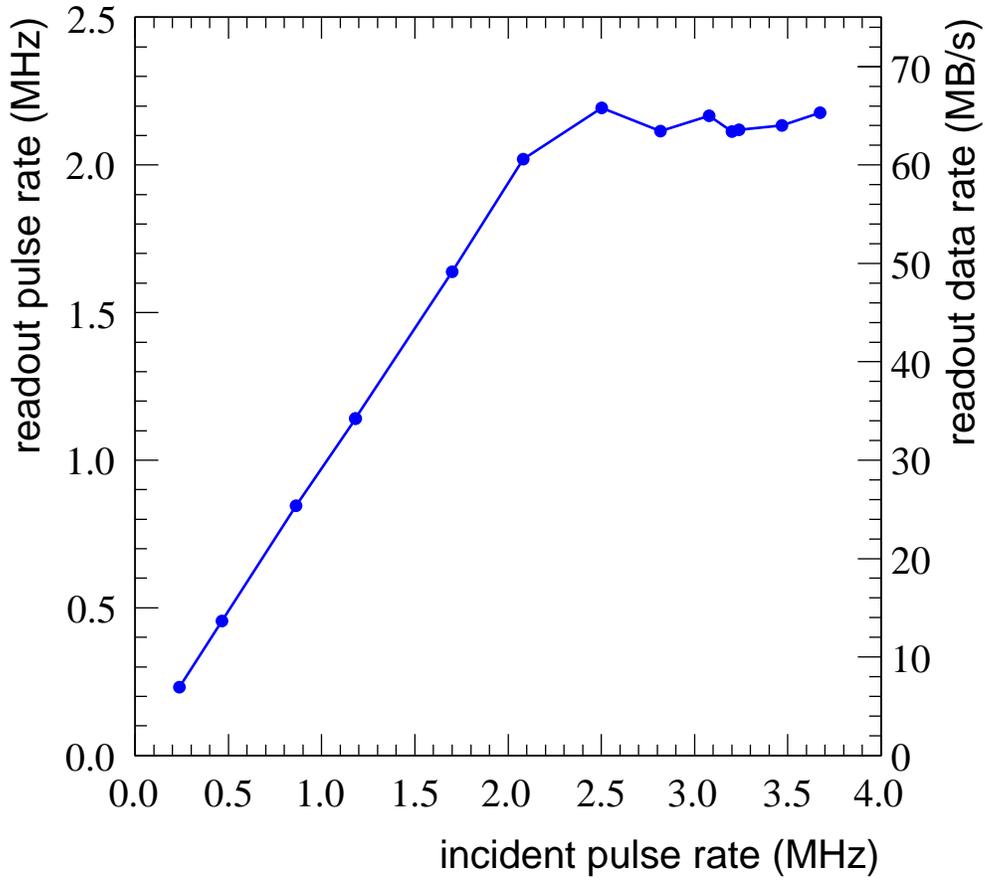,width=0.95\linewidth}}
\end{center}
\caption{The readout pulse rate and readout data rate versus incident pulse rate for the
full data acquisition setup including digitizer read out, lossless compression,
event building and data storage. The maximum throughput of digitized
pulses (decays) is approximately 2~MHz (1~MHz) and is limited by the
data read out and the lossless compression of the digitized pulses.} 
\label{f:performance2}
\end{figure}

In summary, our simple model for the rate performance of the data acquisition
is (i) a fixed deadtime of several percent for
raw data rates below 60~MB/s and (ii) a linearly increasing deadtime 
for greater data rates.
The fixed deadtime was dominated by the read out of the data count 
and fill count registers between the time segments.
The rate-dependent deadtime was dominated by the read out,
lossless compression and databank assembly of the digitized pulses
on the frontend processors.
This interpretation was supported by results obtained from rate tests 
of the data acquisition setup shown in Figs.\ \ref{f:performance1} and \ref{f:performance2}. 
Fig.\ \ref{f:performance1} shows a few percent deadtime below tile pulse rates of 2~MHz
and a linearly increasing deadtime for tile pulse rates above 2~MHz.
Fig.\ \ref{f:performance2} shows the increasing recorded pulse rate 
with increasing incident pulse rate for rates below 2~MHz 
and the saturation of the recorded pulse rate 
with increasing incident pulse rate for rates above 2~MHz.

Note that the limit of 2~MHz (60~MB/s) 
was the rate capacity of the data read out and the lossless compression 
in the waveform digitizer frontend processes. To increase the
capacity one could either increase the number of digitizer
frontends, increase the speed of the frontend processors, or both.

\section{Online monitoring}
\label{s:Monitoring}

\subsection{Online analyzer}
\label{s:Analyzer}

The online analyzer utilized the \textsc{midas} analyzer package 
and a modular, multistage approach to the analysis tasks.
Specifically, the different analysis tasks were implemented 
as individual analyzer modules that could be added or removed 
from the analysis stream as required. 
Each analysis module had access to a global structure
that contained both the raw data from the acquisition read out
({\it e.g.}\ raw data from the digitizer memories)  
and the derived data from the preceding modules
({\it e.g.}\ derived data such as
fill numbers, time stamps, \textsc{adc} arrays).  
Low-level modules were responsible for decompressing the raw data, 
decoding the digitizer data, checking the data integrity,
and filling derived fields in the global structure. 
Histogramming modules were responsible for filling 
the histograms utilized by the low-level data monitoring 
that ensured the experimental setup was operating correctly.
Such histograms included distributions of fill numbers and
hit times for the scintillator tile data and x-y profiles 
and time spectra for the beam monitor data. 
High-level modules were responsible for the ``physics'' analysis
such as fitting of decay curves
by detector position
and monitoring of gains and pedestals by scintillator tile.

\subsection{Run database}
\label{s:Database}

A detailed record of the run-by-run evolution of all relevant quantities
was crucial to ensuring the long term stability of the experimental 
set-up and maintaining a comprehensive record for systematic studies. 
Consequently, an important component of the MuLan data acquisition 
was the automated maintenance of an electronic 
database of the running conditions.

The \textsc{midas} data acquisition package included
support for the \textsc{m}y\textsc{sql} open source database 
-- the \textsc{midas} logger process being responsible
for the transfer of parameters of interest 
from the \textsc{midas} online database to the \textsc{m}y\textsc{sql} database at both the 
start of a run and the end of a run.
The \textsc{m}y\textsc{sql} database contained 
run-by-run information derived from the 
\textsc{midas} online database such as
run start time, run stop time, the number of  
events, and hardware settings including 
the accumulation period and measurement period. 
The run-by-run comments 
that were entered by the shift operators
-- {\it e.g.}\ target material, magnet orientation --
were also copied to the \textsc{m}y\textsc{sql} database by the \textsc{midas} logger.

In addition, a system was developed to enable 
the recording of such physical quantities as the gains, 
pedestals and fitted lifetimes in the scintillator tiles
on a run-by-run basis.
The system was based on a process that read the histogram files
that were generated for the individual runs by the online analyzer,
evaluated the interesting physical quantities such as gains, pedestals
and lifetimes, and wrote the results into the \textsc{m}y\textsc{sql} database.

The database provided a convenient approach
to sorting data according to running conditions
in subsequent analyses.
For example, by using the information recorded 
in the \textsc{m}y\textsc{sql} database we  
automatically maintained summed histograms  
for different running configurations
such as the target material and the magnetic field orientation.

\subsection{Web interface}
\label{s:Interface}

Another important component of the acquisition system
was a custom web interface to the online analyzer histograms, 
slow control data and \textsc{m}y\textsc{sql} database. 
The interface provided a single, simple gateway to 
the online analyzer, slow control and database information
-- irrespective of how the data was derived --
that permitted both local control and monitoring and remote control and monitoring 
of the experiment.
It enabled the plotting of histograms both on a run-by-run basis  
and by experimental configurations such as the target material 
and the magnetic field.
By using a modular structure
the web interface was straightforward for users
to modify or extend.

\begin{figure}
\begin{center} 
\mbox{\epsfig{figure=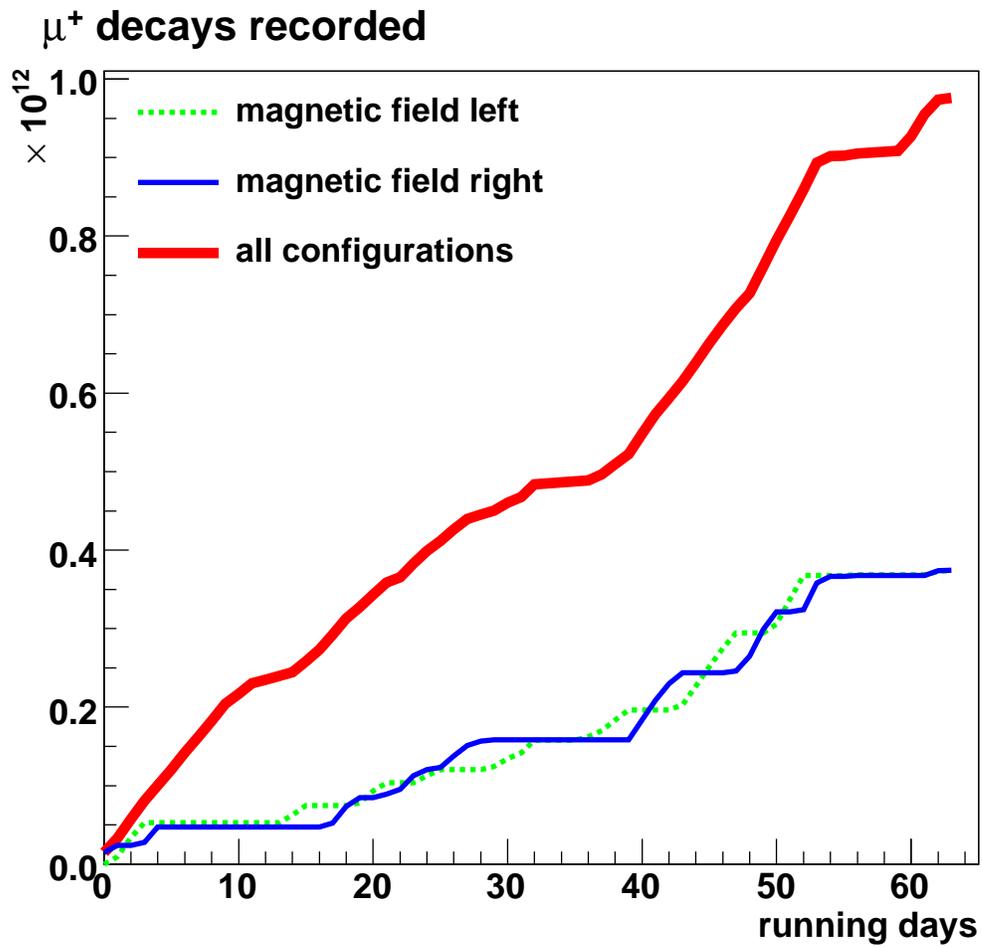,width=0.95\linewidth}}
\end{center}
\caption{Sample trend plot showing the accumulated statistics of muon decays
for different setup configurations (target magnetic field left, target magnetic
field right and all experimental configurations). The 
plot was automatically accumulated using the online analyzer histograms and
the run database information.} 
\label{f:overview}
\end{figure}

\begin{figure}
\begin{center} 
\mbox{\epsfig{figure=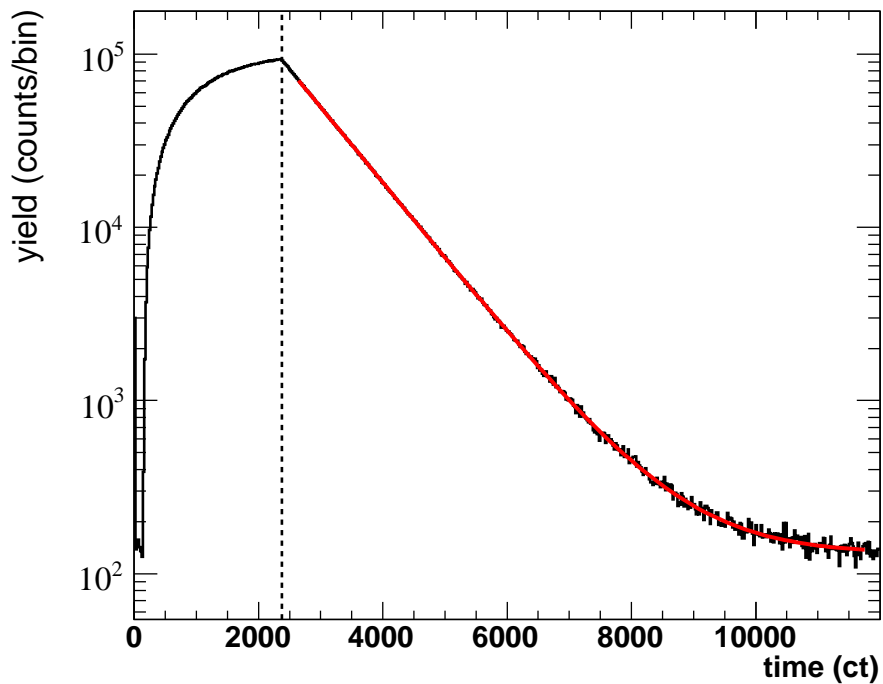,width=0.95\linewidth}}
\end{center}
\caption{Sample run-by-run histogram showing the time distribution of 
tile hits during the fill cycle. The region to the left (right) of 
clock tick 2400 is the accumulation (measurement) period. The solid line
is the least squares fit of the exponential decay curve in the measurement period. The
plot was accumulated using the online analyzer and the fitting results 
are stored in the run database. For this particular run the run gate
was extended into the accumulation period to record the decay positrons 
for the entire fill cycle.} 
\label{f:lifetime}
\end{figure}

The interface enabled plotting of online histograms 
and database information using a dynamic web page 
that utilized a set of \textsc{root} macros and a \textsc{php} language interface 
between the \textsc{root} macros and the \textsc{http} server.
The \textsc{root} macros were responsible for 
generating the histograms and tables 
and creating the output in graphical 
and \textsc{html} formats. The \textsc{php} scripts 
were responsible for processing the user 
requests from the web interface, 
executing the corresponding \textsc{root} macros 
with appropriate input parameters,
and building the \textsc{html}-formatted output web pages. 
Various templates of \textsc{root} macros were 
created to assist users in developing new macros 
that access the information from
the \textsc{m}y\textsc{sql} database, slow control system
and analyzer histograms. 
Fig.\ \ref{f:overview} shows a sample ``trend'' plot 
indicating the accumulation of muon decays
during the 2007 production run.
Fig.\ \ref{f:lifetime} shows a sample run-by-run histogram
indicating the time distribution of tile hits 
during the fill cycle, {\it i.e.}\ both the 
accumulation period and the measurement period.

The web interface also incorporated such 
items as the  run plan and
shift schedules and checklists.  

\section{Summary}
\label{s:Summary}

We have described the data acquisition 
for the MuLan muon lifetime experiment.
The acquisition recorded the digitized output
from scintillator tiles of outgoing positrons 
from muon decay.
The acquisition used
the onboard \textsc{vme} memories in custom-built waveform digitizers,
a repeating cycle of deadtime-free time segments,
and a parallel network of dual-processor machines,
to achieve our goals for integrity, performance and reliability.
The system also featured a custom web interface for 
monitoring experimental conditions that enabled
access to the online histograms, run database
and slow control information.

The system was capable of recording 
muon decays at rates up to approximately 1~MHz 
and read out data at rates up to 
approximately 60~MB/sec with a few percent deadtime. 
The data acquisition system
was used in MuLan production runs in 2006 and 2007
to record a total of $2 \times 10^{12}$ positive 
muon decays.

\section{Acknowledgments}

We would like to thank Stefan Ritt
and Pierre Amaudruz for many valuable communications 
concerning the \textsc{midas} data acquisition package.
This work was supported in part by the
U.S. National Science Foundation.

\newpage

\newpage

\newpage

\newpage

\newpage

\newpage

\newpage

\end{document}